
\input phyzzx

\hsize=6.0truein
\sequentialequations

\overfullrule=0pt
\catcode`\@=11

\def \CO{{\cal O}}

\def\NP{{\it Nucl. Phys.\ }}

\def\PL{{\it Phys. Lett.\ }}

\def\Mod{{\it Mod. Phys. Lett.\ }}

\def\CO{{\cal O}}

\def\eqaligntwo#1{\null\,\vcenter{\openup\jot\m@th
\ialign{\strut\hfil
$\displaystyle{##}$&$\displaystyle{{}##}$&$\displaystyle{{}##}$\hfil
\crcr#1\crcr}}\,}
\catcode`\@=12
\REF\AMP{A. M. Polyakov, \Mod {\bf A2} (1987) 893.}
\REF\KPZ{V. Knizhnik, A. Polyakov and A. Zamolodchikov,
\Mod {\bf A3} (1988) 819.}
\REF\sasha{A. M. Polyakov, ZhETF 63 (1972) 24.}
\REF\bos{A. M. Polyakov, \PL {\bf 103 B} (1981) 207.}
\REF\tsey{E. S. Fradkin and A. A. Tseytlin, \PL {\bf 158 B} (1985) 316.}
\REF\Curt{C. Callan, E. Martinec, M. Perry and D. Friedan,
\NP {\bf B262} (1985) 593.}
\REF\ddk{F.~David, \Mod {\bf A3} (1988)
1651;
J.~Distler and H.~Kawai, \NP {\bf B321} (1989) 509.}

\def\CO{{\cal O}}
\def\eqaligntwo#1{\null\,\vcenter{\openup\jot\m@th
\ialign{\strut\hfil
$\displaystyle{##}$&$\displaystyle{{}##}$&$\displaystyle{{}##}$\hfil
\crcr#1\crcr}}\,}
\catcode`\@=12

\def\qg{quantum gravity}

\nopagenumbers
{\baselineskip=16pt
\line{\hfil PUPT-1421}
\line{\hfil September 1993}
\line{\hfil {\tt hep-th/9309106}}
 }
\title{Gravitational Dressing of Renormalization Group }
\author{I. R. Klebanov,
I. I. Kogan\foot{On leave of absence from ITEP, Moscow, Russia.}
and A. M. Polyakov }
\JHL
\abstract
We study the gravitational dressing of renormalizable
two-dimensional field theories. Our main result is that
the one-loop $\beta$-function is finitely renormalized by
the factor ${k+2\over k+1}$,
where $k$ is the central charge of the gravitational $SL(2, R)$ current
algebra.
\endpage
\pagenumbers

In this letter we report on some minor progress in a very important
subject --- the problem of gravitational dressing. It is well known that
string theory in any dimension is described by two-dimensional field
theory coupled to two-dimensional \qg. This coupling leads to
the physically important change of the scaling dimensions and to
the appearance of unexpected symmetries [\AMP, \KPZ]. There are many important
problems in string theory where one has to study the gravitational
dressing of two-dimensional models which are renormalizable, but
not conformally invariant. Here we will focus on precisely this
new situation. More concretely, we will show how the renormalization
group equation is deformed (in the one-loop approximation) by the
gravitational dressing.

The Gell-Mann--Low $\beta$-function appears when one perturbs a
conformal field theory by some marginal operators
$$ S= S_0 +\sum_n \lambda_n \int O_n (x) d^2 x
\eqn\pert $$
The renormalization group equation for the coupling constants
$\lambda_n$ has the form
$$ {d \lambda_n\over d\log \Lambda } =\beta_n (\lambda)
$$
where $\Lambda$ is the ultraviolet cut-off.

It was realized long ago [\sasha] that the Gell-Mann--Low $\beta$-function
is related to the structure constants of the operator algebra.
Namely, if we have an operator product expansion (OPE)
$$ O_n (x) O_m (0) ={1\over |x|^2} g^{nm}_l O_l (0) +\ldots
$$
then
$$\beta_l (\lambda) =2\pi g^{nm}_l \lambda_n \lambda_m +\CO(\lambda^3)
$$
In order to find these structure constants in any theory, it is
sufficient to calculate the 2-point functions and the
3-point functions of the perturbing operators. Below we will
perform this calculation in the novel situation where the
two-dimensional \qg\ is ``turned on''. Let us explain the precise meaning of
this. In order to describe the effects of gravity it is
advantageous to use the light-cone gauge of ref. [\AMP], since the
coordinate cut-off in this gauge is equivalent to the physical cut-off.
The line element is
$ds^2 = 2 dx^+ dx^- + h_{++}(x) (dx^+)^2$, so that the gravitation is
described by a field $ h_{++}(x) $ coupled to the energy momentum
tensor. Hence, ``turning the gravity on'' means replacing eq. \pert\
with the action
$$ S= S_0 +\sum_n \lambda_n \int O_n (x) d^2 x
+\int h_{++}(x) T_{--} (x) d^2 x
\ ,\eqn\dressac$$
where $T_{--} (x)$ is a component of the matter energy momentum
tensor, and $h_{++}(x)$ is to be treated as a quantum field.

Without losing much generality, we will concentrate on the case
where the marginal operators $O_n$ can be represented as
products of right-moving and left-moving conserved currents.
Thus, we chose the second term in \dressac\ to be of the form
$$ \sum_{A, A'} \lambda_{A A'} \int d^2 x~ J_-^A (x^-)
J_+^{A'} (x^+)\ .
$$
The 2-point and 3-point functions of the currents are
$$\eqalign{&<J_-^A(x_1) J_-^B (x_2)>=  K\delta^{AB}
{1\over (x_1^- -x_2^-)^2}\ , \cr
&<J_-^A(x_1) J_-^B (x_2) J_-^C (x_3)>=  f^{ABC}
{1\over (x_1^- -x_2^-) (x_2^- -x_3^- ) (x_3^- -x_1^-)} \ ,\cr }
\eqn\pro $$
and similarly for $J_+^{A'}$.
In this case the structure constants of the OPE are the products
of the structure constants of the current algebra,
$$ g^{(B B')~(C C')}_{(A A')} = f_A^{BC} f_{A'}^{B' C' }\ ,
\eqn\sc$$
where $f_A^{BC}={1\over K}\delta_{AD}f^{DBC}$.

Now, let us consider the effects of the gravitational dressing.
In the light-cone gauge, the correlation functions of the $J_+^{A'}$ are
unchanged by gravity, while those of the $J_-^A$ are renormalized,
so that the first factor in eq. \sc\ changes. In order to calculate this
effect, we need to perform the following functional integral,
$$\ll J_-^{A_1}(x_1) \ldots J_-^{A_n}(x_n) \gg~=
\int [D h_{++}(x)] < J_-^{A_1}(x_1) \ldots J_-^{A_n}(x_n)
e^{i\int h_{++}(x) T_{--} (x) d^2 x}>
\eqn\dresscorr$$
where the double brackets denote the gravitationally dressed
correlators. This path integral is calculable essentially because
the Ward identities determine the correlation functions containing
any number of insertions of $T_{--}$. As a result, the dressed
correlator \dresscorr\ satisfies a rather peculiar differential equation of
the type derived in ref. [\AMP]. This equation is derived from the
conservation law for the dressed currents,
$$\partial_+ J^A_-  =\partial_- (h_{++}J^A_-)
\ ,\eqn\cont$$
which comes from the gauge-fixed version of $D^\beta J^A_\beta=0$.
The product $h_{++}J^A_-$ has to be properly regularized.
It can be calculated explicitly because the Ward identities
determine the correlation functions of $h_{++}$. In particular,
we have [\AMP]
$$\eqalign{&
\ll h_{++}(w) J^{B_1} (x_1) J^{B_2}(x_2) \ldots J^{B_n}(x_n) \gg~ =\cr
&-{1\over k+2}\sum_{i=1}^n {\partial\over \partial x_i^-}
{(w^-- x_i^-)^2\over w^+- x_i^+}
\ll J^{B_1} (x_1) J^{B_2}(x_2) \ldots J^{B_n}(x_n) \gg\ , \cr }
\eqn\gW$$
where $k$ is the central charge of the gravitational $SL(2, R)$ current
algebra, related to the matter central charge by the formula [\KPZ]
$$ k+2 ={1\over 12}\left (c-13 \pm \sqrt{(c-1)(c-25)}\right )
\eqn\central$$
In order to match onto the semiclassical limits $|c|\to \infty$,
the $+$ sign is chosen for $c\geq 25$, and the $-$ sign
for $c\leq 1$. Combining eqs. \cont\ and \gW, we obtain the following
differential equation,
$$\eqalign{&{\partial\over \partial w^+}
\ll J_-^A(w) J_-^{B_1} (x_1) J_-^{B_2}(x_2) \ldots J_-^{B_n}(x_n) \gg~ =\cr
&-{1\over k+2}\sum_{i=1}^n {\partial^2\over \partial w^-\partial x_i^-}
{(w^-- x_i^-)^2\over w^+- x_i^+}
\ll J_-^A(w) J_-^{B_1} (x_1) J_-^{B_2}(x_2) \ldots J_-^{B_n}(x_n) \gg \cr
& -2\pi i \sum_{i=1}^n
{\partial\over \partial w^-}\delta^2(w-x_i)
K \delta^{A {B_i}} \ll J_-^{B_1}(x_1)\ldots J_-^{B_{i-1}}(x_{i-1})
J_-^{B_{i+1}}(x_{i+1})\ldots J_-^{B_n}(x_n) \gg \cr
& +2\pi i \sum_{i=1}^n \delta^2(w-x_i)
f_{C}^{A {B_i}}
\ll J_-^{B_1}(x_1)\ldots J_-^{B_{i-1}}(x_{i-1}) J_-^C (x_i)
J_-^{B_{i+1}}(x_{i+1})\ldots J_-^{B_n}(x_n) \gg
\ ,\cr }
\eqn\full$$
where $K$ is the central charge of the flat space current algebra,
and $f^{ABC}$ are its structure constants from eq. \pro.
Note that the first term
on the right-hand side of eq. \full\ can be regarded as the gravitational
violation of the right-moving nature of the current, while the remaining
delta-function terms are the violations that take place even without
the coupling to gravity. It is an interesting exercise to check
eq. \full\ in the $1/c$ expansion directly from
the path integral \dresscorr.

Eqs. \full\ in principle solve the problem of gravitational dressing.
We suspect that they contain many mathematical and physical surprises.
Presently we will make only a limited use of them. Namely, we determine
the 2-point and the 3-point functions, and thus solve the problem of
gravitational dressing for the one-loop renormalization group
$\beta$-function. The key point is that the $x$-dependence in these two
cases is completely fixed by the projective invariance,
$$\eqalign{&\ll J_-^A(x_1) J_-^B (x_2)\gg~= \tilde K\delta^{AB}
{1\over (x_1^- -x_2^-)^2}\ , \cr
&\ll J_-^A(x_1) J_-^B (x_2) J_-^C (x_3)\gg~= \tilde f^{ABC}
{1\over (x_1^- -x_2^-) (x_2^- -x_3^- ) (x_3^- -x_1^-)} \ .\cr }
\eqn\project $$
In order to determine the constants $\tilde K$ and $\tilde f^{ABC}$,
we substitute the ansatz \project\ into eqs. \full. From the equation
for the 2-point function, we find
$$\tilde K = K -{1\over k+2} \tilde K \eqn\relo$$
{}From the equation for the 3-point function, we obtain a relation
$$ \tilde f^{ABC}={\tilde K\over K} f^{ABC}+{1\over k+2}\tilde f^{ABC}
\eqn\relt$$
Solving eqs. \relo\ and \relt, we find
$$\eqalign{&\tilde K={k+2\over k+3} K\ ,\cr &
\tilde f^{ABC}=f^{ABC} {(k+2)^2\over (k+1)(k+3)} \ .\cr }
\eqn\main $$
Since the indices are now raised and lowered with the metric given
by the two-point function in eq. \project, we have
$$ \tilde f^{BC}_A={1\over \tilde K} \delta_{AD} \tilde f^{DBC}
={1\over K} \delta_{AD}f^{DBC} {k+2\over k+1}
$$
Recalling that the structure constants are given by eq. \sc, we
have
$$ \tilde g^{(B B')~(C C')}_{(A A')} =
{k+2\over k+1} g^{(B B')~(C C')}_{(A A')}
\eqn\mainres$$
\ie\ the one-loop $\beta$-function is multiplicatively renormalized by
the gravitational dressing. This
universal ``renormalization of the renormalization group''
constitutes our main result.

In order to further investigate the gravitational dressing of the
renormalization group flow, we need to calculate the higher-order
terms in the $\beta$-function. The first such term, which is
$\CO(\lambda^3)$, is determined by the 4-point function of the
currents. Here the calculation is more complicated because
the $x$-dependence is not completely fixed. Instead, the projective
invariance requires that
$$\ll J_-^A(w) J_-^B (z) J_-^C(x) J_-^D (y)\gg~  =
{F^{ABCD}(t, \bar t)\over (x-y)^2 (z-w)^2}
\ ,\eqn\repre$$
where\foot{Here we find it convenient to continue to the Euclidean
signature so that $x^-$ is replaced by a complex variable $x$,
and $x^+$ -- by its complex conjugate $\bar x$.}
$$ t={(w-x)(z-y)\over (w-y)(z-x)}
$$
After substituting this form into eq. \full\ for the 4-point function,
we find
$$-(k+2)\bar t~{\partial F\over\partial\bar t}=
2F {t+1\over t-1}-3t~{\partial F\over\partial t}
+(\bar t-t) {t-1\over \bar t-1}~{\partial \over\partial t}
\left (t~{\partial F\over\partial t}\right )
\eqn\rich$$
The boundary conditions on $F$ at the special points
$t=0,~ 1$ and $\infty$ are determined by the singular terms in
eq. \full. Eqn. \rich\
has a rich variety of solutions.
Perturbation theory in $1/c$ suggests that there
is only one physically acceptable solution.
Let us consider, for instance, a theory of $c$ free massless Dirac
fermions, where the currents are represented as
$J_-^A= \psi_{-j}^\dagger \lambda^A_{ji}\psi_{-i}$,
and $\lambda^A$ are generators of $SU(N)$. Here the 4-point function
is of the form \repre, and $F^{ABCD}(t, \bar t)$ is a sum of terms
distinguished by their $SU(N)$ structure. There are two types of
contributions:
$$\delta^{AB}\delta^{CD}\left (1- {12\over c}~{t+1\over t-1}\log |t|^2
+\CO(1/c^2)\right )
$$
plus two other $\delta\cdot\delta$ terms, and
$$\Tr (\lambda^A\lambda^B\lambda^C\lambda^D)
\left [\left (1+{12\over c}\right ){t-1\over t}-{6\over c}
\left (\log |t|^2+{(t-1)^2\over t^2}\log |t-1|^2\right )+\CO(1/c^2)\right ]
$$
plus five other terms of this type. The correct solution should be
chosen to match these perturbative results.
Let us also note that the acquired dependence of the 4-point function on
$\bar t$ is a purely gravitational effect that is probably connected
with the appearance of new states, the gravitational descendants.

We have found our main result, eq. \mainres, using the light-cone gauge
for the two-dimensional \qg. We believe, however, that this result
is completely universal and gauge independent. In order to illustrate
how a similar calculation proceeds in the conformal gauge, we will
study the gravitational dressing of the $O(N)$ model.
The model is defined by the action
$$ S={1\over 2 e^2}\int d^2 x\sqrt g g^{\mu\nu}\partial_\mu \vec n
\cdot \partial_\nu \vec n
\ ,$$
where $\vec n$ is an $N$-component vector of length 1.
If we choose the conformal gauge [\bos]
$g_{\mu\nu}=e^{-\phi}\delta_{\mu\nu}$,
then $\phi$ becomes dynamical and
can be regarded as an extra coordinate of critical string
theory. We will assume that the target space metric in this theory
is of the form
$$ ds^2 = d\phi^2+a^2 (\phi) d\Omega^2
\ ,
$$
where $d\Omega^2$ is the metric of a unit sphere in $N$ dimensions.
We also need to introduce the dilaton field $\Phi(\phi)$, which couples
to the world sheet curvature [\tsey].
The variable $\phi$ defines the world sheet scale,
and the effective radius of the sphere becomes scale-dependent through
quantum effects. The form of this dependence is determined by
the conformal invariance or, equivalently, by the consistency of the
critical string. In the one-loop approximation, the consistency
equations are well-known to be [\Curt]
$$\eqalign{& R_{ij} =\nabla_i \nabla_j \Phi\ ,\cr
& \nabla^2\Phi +(\nabla\Phi)^2 = {26-N\over 3}\ .\cr
}\eqn\consist $$
Substituting our ansatz for $G_{ij}$ and $\Phi$ into these equations,
we arrive at
$$\eqalign{& N-2 -aa''- (N-2)(a')^2= aa'\Phi'\ ,\cr
&\Phi''+ (N-1){a''\over a}=0\ ,\cr
&\Phi''+ (N-1){a'\over a}\Phi'+(\Phi')^2={26-N\over 3}\ .\cr
} $$
For large $\phi$, the asymptotic form of the needed
solution to these equations is
$$\eqalign{&\Phi= Q\phi +\CO(\log\phi)\ ,\cr
&a^2=2{N-2\over Q}\phi +\CO(\log\phi)\ ,\cr }
$$
where $Q=\sqrt{25-c\over 3}$ and $c=N-1$.
In order to calculate the gravitationally dressed $\beta$-function,
we need a precise definition of the physical scale. A sensible
definition is to identify the scale $\Lambda^{-2}$
with the cosmological constant
operator, also known as the ``tachyon background'' $T(\phi)$.
\foot{A good test of this definition is the eventual
agreement of the one-loop $\beta$-function with the
light-cone gauge calculation. It is possible, however, that
the correct definition of scale is more subtle for higher-order
calculations in the conformal gauge.}
Its leading asymptotic for large $\phi$ is [\ddk]
$$\eqalign{&\Lambda^{-2}\sim T(\phi)\sim e^{-\alpha\phi}\ , \cr
&\alpha={\sqrt{25-c}-\sqrt{1-c}\over \sqrt {12}}\ . \cr }
$$
Thus, the gravitationally dressed renormalization group equation is
$${ d a^2\over d\log\Lambda} =2(N-2) {2\over Q\alpha}+\CO(a^{-2})
\ .$$
The form of the equation before coupling to gravity is well known to be
$${ d a^2\over d\log\Lambda} =2(N-2) +\CO(a^{-2})
\ ,$$
where $a^2=4\pi/e^2$.
Therefore, the gravitational renormalization factor for the one-loop
$\beta$-function is ${2\over Q\alpha}$ which, with eq. \central,
easily reduces to ${k+2\over k+1}$. This provides a non-trivial
check of the universality of our results.

We suspect that there are similar universal gravitational
renormalizations for the sub-leading terms in the $\beta$-function.
The problem of determination of the full gravitational effect on the
renormalization group is a fascinating challenge.

\ack

The research of I. R. K. is supported in part by
DOE grant DE-AC02-76WRO3072, NSF Presidential Young Investigator
Award PHY-9157482, James S. McDonnell Foundation grant No. 91-48, and
an A.P. Sloan Foundation Research Fellowship. The research of A. M. P.
and I. I. K. is supported in part by the NSF grant
PHY90-21984.

\singlespace
\refout

\bye